\documentclass[11pt, a4paper,  notitlepage]{extarticle}

\usepackage[left=2.5cm, right=2.5cm, top=3cm, bottom=3cm, bindingoffset=0cm]{geometry}
\usepackage{graphicx}
\usepackage{mathtools}
\usepackage[affil-it]{authblk}
\usepackage{url}
\usepackage{amsmath}
\usepackage{amssymb}
\usepackage{multicol}
\usepackage{float}
\usepackage[section]{placeins}
\usepackage{hyperref}
\usepackage[title]{appendix}

\begin{document}
\sloppy
\title{AI Approaches to Qualitative and Quantitative News Analytics on NATO Unity}
\author{Bohdan M.  Pavlyshenko \\  \small{Ivan Franko National University of Lviv,  Ukraine \\ b.pavlyshenko@gmail.com,  www.linkedin.com/in/bpavlyshenko/ }}
\maketitle

\begin{abstract}
The paper considers the use of GPT models with retrieval-augmented generation (RAG) for qualitative and quantitative analytics on NATO sentiments, NATO unity and NATO Article 5 trust opinion scores in different web sources:  news sites found via Google Search API, Youtube videos with comments, and Reddit discussions.  A RAG approach using GPT-4.1 model  was applied to analyse  news where NATO related topics were discussed.   
Two levels of RAG analytics were used: on the first level, the GPT model generates qualitative news 
summaries and quantitative opinion scores using zero-shot prompts; on the second level, the GPT model generates the summary of news summaries. 
Quantitative news opinion scores generated by the GPT model were analysed using Bayesian regression to get trend lines. The distributions found for the regression parameters make it possible to  analyse an uncertainty in specified news opinion score trends.
Obtained results show a downward trend for analysed scores of opinion related to  NATO unity.
 This approach does not aim to conduct real political analysis; rather, it consider AI based approaches which can be used for further analytics 
 as a part of a complex analytical approach. The obtained results demonstrate that the use of GPT models for news analysis  can give informative qualitative and quantitative analytics, providing important insights. 
 The dynamic model based on neural ordinary differential equations was considered for modelling public opinions. This approach  makes it possible to  analyse different scenarios for evolving public opinions.

\end{abstract}

\tableofcontents
\section{Introduction}
Using AI approaches for analysing news text information is a highly promising method of understanding key processes in society, especially public opinions on different points of social and political life. One of such points is the public opinion on NATO unity and security. Currently, during the active phase of Russian-Ukrainian war, it is important to understand all possible threats.

One of the most important parts of NATO agreement is Article 5.
Article 5 of NATO Agreement states that if a NATO country is the victim of an armed attack, each Alliance member will consider this act of violence as an armed attack against all members and will take such measures as it deems necessary, "including the use of armed force," to come to the aid of the attacked country.
NATO unity is a crucial aspect of the alliance's strength, enabling it to effectively address collective defense and maintain peace and security. This unity is demonstrated through various means, including joint military exercises, shared values, and cooperation among member states. 
NATO is built on the foundation of shared values, including freedom, democracy, and the rule of law, which foster a sense of common purpose and strengthen the alliance. 
Member states work together to support each other, both militarily and politically, in times of crisis or when addressing common challenges. This cooperation can range from military deployments to diplomatic efforts, ensuring a coordinated response to threats. 
 Different aspects related to NATO unity are considered in\cite{tardy2022nato, kramer2009myth, lipka2019london}.

At the same time, NATO unity faces different challenges. 
 While the alliance generally operates on a consensus-based system, individual member states may have differing views on certain issues, which can create tensions and challenge unity. The war in Ukraine and other global challenges can strain relations within NATO, requiring the alliance to find ways to maintain unity in the face of adversity. 
NATO unity is both a military doctrine and a political challenge.
Article 5 binds the alliance together, but the real test lies in political will -- whether diverse democracies can respond together when facing emerging threats.
Main challenges of NATO are widely discussed in news media and social networks. So, it is important to find qualitative and quantitative 
of public opinion on NATO unity.  One must understand public and news opinion on Article 5 spirit, if it will work 
in case of real threats and military escalation. In our previous works~\cite{pavlyshenko2022methods}, we considered different approaches for detecting manipulative and fake news. 
A deep learning approach for fake news detection has been analyzed. The use
of the theory of frequent itemsets and association rules, graph theory for news
trends analytics have been considered.
In~\cite{pavlyshenko2023analysis} we consider the possibility of fine-tuning Llama 2 large language model (LLM) for the disinformation analysis and fake news detection.  For fine-tuning, the PEFT/LoRA based approach was used. In the study, 
the model was fine-tuned for the following tasks: analysing a text on revealing disinformation and  propaganda narratives, fact checking, fake news detection,  manipulation analytics, extracting named entities with their sentiments.
The obtained results show that the fine-tuned Llama 2 model can perform a deep analysis of texts and reveal complex styles and narratives.

The main goal of this work is to consider AI based approaches for qualitative and quantitative analytics of NATO unity and public trust that Article 5 will effectively work when political and military threats come into reality. 
We do not concentrate efforts analysis of  social and political aspects of NATO unity, but rather on considering AI approaches to conduct such kind of analytics. 

\FloatBarrier
\section{Using GPT Models for News Opinion Analytics }
\subsection{Analytics of Web Resources Found by Google API}
\begin{figure}[H]
\center
\includegraphics[width=1\linewidth]{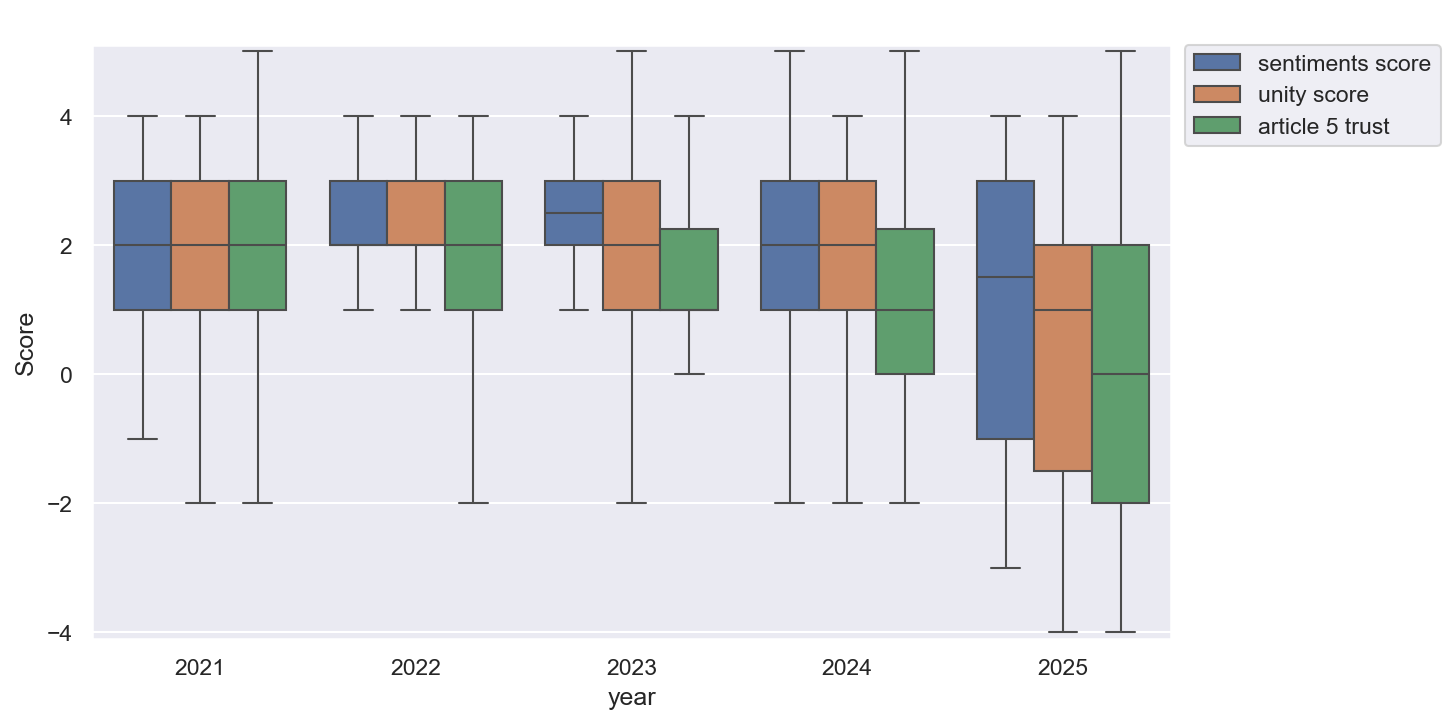}
\caption{Box plot for quantitative scores}
\label{boxplots_scores_all}
\end{figure}

\begin{figure}[H]
\center
\includegraphics[width=0.7\linewidth]{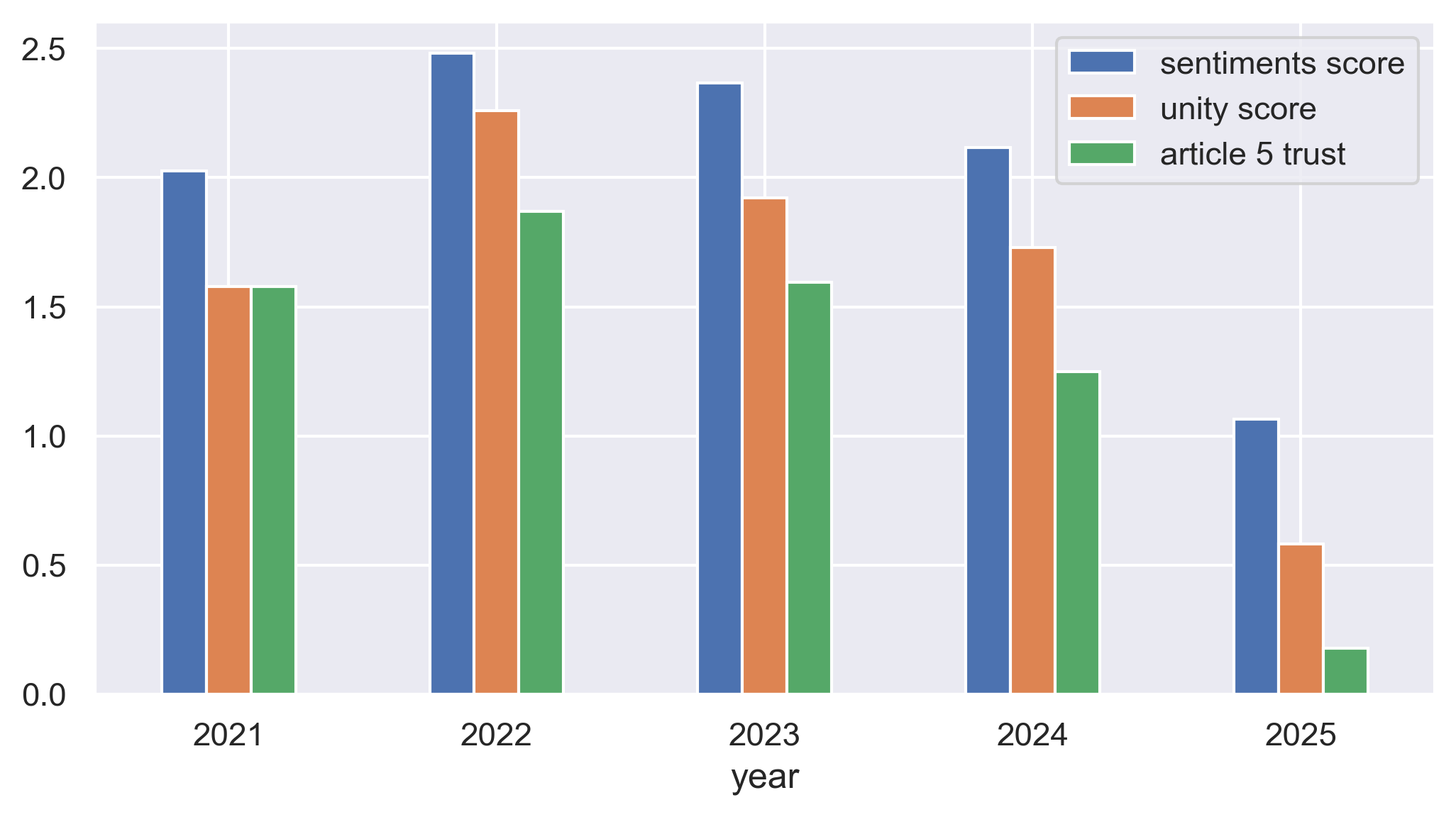}
\caption{Mean values for quantitative scores}
\label{scores_mean_all}
\end{figure}

News data for the analysis were received from web resources in two steps:
\begin{enumerate}
\item Retrieiving URLs for relevant web resources using \textit{Google Search API}. For this purpose, Python library \textit{Google API Clent} was used.
\item Extracting information from web resources given their URLs. For this purpose \textit{SeleniumURLLoader} from 
\textit{LangChain} python library~\cite{langchaingithub} was used. 
\end{enumerate}
 The data from each extracted resource were analyzed using Open AI with GPT-4.1 model. 
The text data for the analysis were included in a created prompt for the  GPT-4.1 model, where we specified the options for the analysis, e.g. 
what and how to summarize, which points to highlight, what quantitative scores should be generated. 
We also specified the instructions for the output format which should be in JSON with an appropriate structure. 
The instructions specified the generation of opinion scores.
In the prompt, the instructions  to summarize key points of news texts and to find quantitative scores for sentiments towards NATO, NATO unity scores and score for Article 5 trust were given. Scoring instructions specified to calculate scores in the range [-5, 5], where -5 means max negative and 5 means max positive score for the characteristics under analysis.
We used the following search queries for Google Search API: 'NATO unity OR security', 'NATO article 5' separately for each annual period within 2021-2025, up to April 10, 2025. For each query and annual period, top 70 web resources were chosen.   Some web resource were not loaded due to site restrictions.
 As a result of GPT analytics, 258 resources where selected as relevant.  
Fig.~\ref{boxplots_scores_all} shows box plot for quantitative scores. 
Fig.~\ref{scores_mean_all} shows mean values for quantitative scores.
\FloatBarrier

On the second level of GPT analytics, we created a prompt for summarizing  summaries of news which were obtained on the first level using GPT-4.1 API. The following resulting summaries have been received:
{\fontsize{9}{11}\selectfont \begin{verbatim}

**Summary of Summaries: NATO Unity, Security, and Article 5 Trust (2021-2025)**

**Challenges for NATO Unity, Security, and Trust in Article 5**

From 2021 through early 2025, the core themes emerging from a comprehensive review of news and
analytical resources are persistent and, in some cases, intensifying challenges to NATO unity,
collective security, and trust in Article 5. These challenges arise from:

- **Divergent Threat Perceptions:** Members continue to disagree on prioritizing threats, notably on
Russia vs. China, hybrid vs. traditional warfare, and security interests in regions like the Arctic,
Black Sea, and Indo-Pacific. While Russia’s aggression has rallied much of the alliance, questions
remain over extending unity to new domains and geographies (cyber, space, economic security).

- **Political and Policy Differences:** There is ongoing friction over defense spending, burden-
sharing, enlargement (notably concerning Ukraine, Sweden, Finland), and differing national
approaches to security (illustrated by debates over climate, human security, and defense industrial
policies).

- **US Commitment Doubts:** The reliability of the US — NATO’s central pillar — is under historic
scrutiny, especially with shifting presidential rhetoric (notably from Trump), Congressional
assertions of war powers, and threats to condition or withhold U.S. support for allies not meeting
spending targets. This has prompted increased calls for European defense autonomy and alternative
security arrangements.

- **Consensus-based Decision-making:** The need for unanimity in NATO decision-making means that
individual states can block collective actions, introducing veto risks in crises and undermining the
perceived automaticity of mutual defense.

- **Hybrid and Non-traditional Threats:** Rising cyberattacks, disinformation, sabotage, and other
hybrid tactics test NATO’s ability to determine when and how to respond collectively. There is
persistent ambiguity about whether— and how— Article 5 would apply in such cases.

- **Enlargement and External Pressure:** Prospective or partial security guarantees for Ukraine and
ongoing disputes over extending full Article 5 protections to new members (or specific territories)
risk exposing alliance divisions. Russian and, to a lesser degree, Chinese hybrid activities and
direct threats are actively testing NATO’s internal cohesion and resolve.

**Time Trends: Unity, Sentiment, and Article 5 Trust Scores**

- **Unity Scores:**
  - *2021-2022:* Despite a surge in rhetorical and symbolic unity around Russia’s invasion of
Ukraine (unity scores often 3-4 in summit-related documents and high-level statements), operational
and political divisions persisted, especially around enlargement, defense spending, and non-
traditional threats. Lap points: partial unity or visible strain, especially in critical moments
such as the Afghanistan withdrawal, debates around Ukraine’s membership, and cyber crises.
  - *2023-2024:* Unity shows brief, event-driven peaks (e.g., Finland and Sweden’s accession, NATO’s
anniversaries) but is overall more fragile than in earlier years, as reflected in a greater number
of resources with unity scores 0-2 or even negative. Unity declines further in late 2024 and 2025,
notably with U.S. political shifts, open threats to alliance guarantees, and European moves toward
strategic alternatives or autonomous defense.
  - *2025:* Sharp declines in unity scores are common in resources discussing new US administration
attitudes, hypothetical intra-alliance conflicts, and greater public questioning of the alliance’s
purpose and methods, especially surrounding Trump-linked rhetoric.

- **Sentiment Scores:**
  - *2021-mid-2022:* Generally positive (2-4) amid public reaffirmations and united responses to
immediate threats. Documents from defense/public fora are optimistic, though more analytical sources
flag underlying problems.
  - *Late 2022-2025:* Sentiment erodes, with rising negative or neutral scores as unity, operational
cooperation, and trust are undermined by real or anticipated US disengagement, overtly transactional
rhetoric, and unresolved intra-European disagreements.

- **Article 5 Trust Scores:**
  - *2021:* Mixed. Scores from -1 to 5, indicating that while summit rhetoric was strong, deeper
analysis shows persistent doubts about automaticity, scope (e.g., cyber, hybrid, out-of-area), and
willingness to act (as in the Baltic, Black Sea, or Indo-Pacific). The withdrawal from Afghanistan
and pressure on spending worsened trust among some EU members.
  - *2022:* Initially boosted by rhetorical reaffirmations post-Ukraine invasion (scores up to 4-5
in official pronouncements), but tested by ambiguous crises (missile incidents, cyber events), the
practicalities of invoking Article 5, and case-by-case ambiguities (especially for new or non-
European members and hybrid attacks).
  - *2023-2024:* Declining trend, with trust scores between 0 and 2 predominant except for formal
summit documents. Explicit Trump-era and Congressional debates (along with European consideration of
autonomy) further erode trust, reflected in increasing skepticism in frontline states and even
traditional NATO stalwarts.
  - *2025:* Heightened doubts, with frequent -2 to -4 trust scores in analyses, reflecting deep
anxiety about U.S. reliability, more hypothetical or conditional security guarantees (e.g., to parts
of Ukraine), and discussions about alternative European defense prospects.

**Groups of News Sites by Opinion on NATO Unity and Article 5 Trust**

- **Pro-Unity, High Article 5 Trust:**
  - *Government, military, and NATO-linked sources* (army.mil, defense.gov, af.mil, navy.mil,
veterans.gc.ca, europa.eu, NATO/EU official sites) repeatedly affirm strong unity and robust Article
5 trust, especially during summits, anniversaries, or after successful exercises/integration events.
  - *Allied national foreign ministries or defense sites* (auswaertiges-amt.de, government.nl)
consistently reinforce positive sentiment and trust in collective defense.

- **Supportive but Cautiously Analytical:**
  - *Policy think tanks, mainstream U.S. and European news organizations, and defense-focused
academic outlets* (csis.org, atlanticcouncil.org, rusi.org, carnegieendowment.org, SWP-Berlin,
GMFUS, BBC, NYTimes) typically score unity and Article 5 trust as moderately high but frequently
point to operational, political, and strategic obstacles, especially around burden-sharing, U.S.
commitment, and new threat domains.

- **Skeptical/Critical; Highlighting Divisions and Doubts:**
  - *Academic, specialist, and critical international sites* (cato.org, lawfaremedia.org, quora.com,
reason.com, responsiblestatecraft.org, westpoint.edu, house.gov/senate.gov in Congressional context,
various op-eds) consistently foreground the legal, political, and operational caveats that challenge
unity and trust. They express concern about U.S. conditionality, the lack of automatism, and the
risk of failing to uphold Article 5 under political or legal duress.
  - *Some mainstream news sites* (BBC, CNN, businessinsider.com, theguardian.com, bloomberg.com,
politico.eu) register discouraging signals when reporting on U.S. or major-member leaders’ overt
questioning of Article 5, or open intra-alliance discord.

- **Outright Negative or Alarmist:**
  - *Cato Institute, some U.S./UK/EU analysts, and critical think tanks* (cato.org,
verfassungsblog.de, responsiblestatecraft.org, msnbc.com, nbcnews.com) are unsparing in their
warnings, documenting a collapse or acute risk to unity and trust, especially with Trump-linked
rhetoric, rising calls for European autonomy, and hypothetical scenarios involving intra-alliance
conflict, green-lighting of Russian actions, or abandonment.
  - *International and external observer commentary and opposition-leaning sources*
(kyivindependent.com, pravda.com.ua, aa.com.tr, lrt.lt) flag deep skepticism about the extension or
flexibility of Article 5, especially regarding Ukraine, and the danger posed by informal and
ambiguous guarantees.

**Conclusion**

While the 2021-2022 period saw bursts of unity and reaffirmed trust in Article 5 under duress
(especially after the Russian invasion of Ukraine), the subsequent years reveal mounting and
unresolved doubts—exacerbated by political rhetoric in leading member states, internal disagreements
on enlargement, hybrid threats, and the lack of consensus or clarity on collective defense’s scope
and implementation.

- **Unity is profoundly event- and context-dependent**: It peaks at summits or in collective
exercises, but strategic, legal, and political cracks are increasingly visible, especially as more
actors question American leadership or propose alternative security arrangements.
- **Sentiment and trust are declining**: The balance of analytical and official resources shows a
shift from cautious optimism to apprehension or alarm, particularly amid U.S. political
oscillations, hybrid threats, and debates over the automaticity and scope of article 5 protections.
- **Groups differ by tone and assessment**: Official, government-linked, and mainstream policy sites
are more optimistic, but policy critics and think tanks increasingly warn about procedural
ambiguities, wavering political will, and the rising risk of alliance fracture in a crisis.

**In sum:** NATO remains resilient by many operational metrics, but faces its most acute test of
internal unity and external credibility since its founding, centered on whether political will,
legal clarity, and strategic solidarity can withstand new and old threats—and especially whether
Article 5 will remain an automatic, dependable guarantee for all members.
\end{verbatim}
}

We also conducted the research on analytics with natural language queries. We tested the GPT-4.1 prompt with inserted
natural language query ”Give top 10 web resources with links where Article 5’s credibility is
openly doubted”. The results of GPT response are presented in the Appendix. 

We conducted a similar analysis for specified web resources: \textit{cnn.com,  nbcnews.com, nytimes.com, foxnews.com}.
Obtained qualitative results are presented in the Appendix. Fig.~\ref{google_sites_img} shows time trends of quantitative scores for these specified resources.
 One can see that the dynamics of quantitative scores for NATO unity and 
Article 5 trust have a downwards trend. 

\FloatBarrier
\begin{figure}[H]
\center
\includegraphics[width=0.85\linewidth]{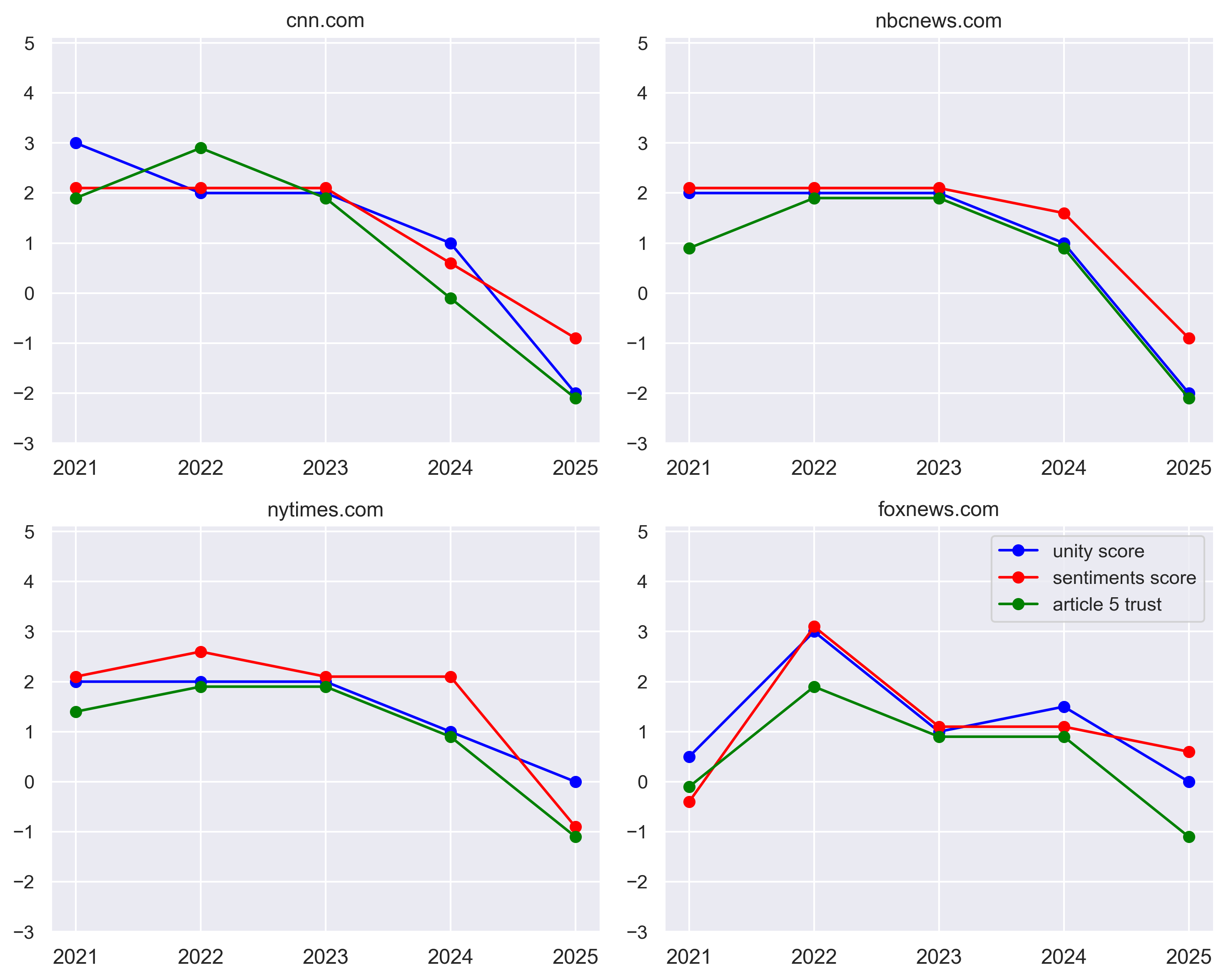}
\caption{Quantitative opinion scores in news for web resources}
\label{google_sites_img}
\end{figure}

\FloatBarrier
\subsection{Youtube Videos Analytics}
\begin{figure}[H]
\center
\includegraphics[width=0.85\linewidth]{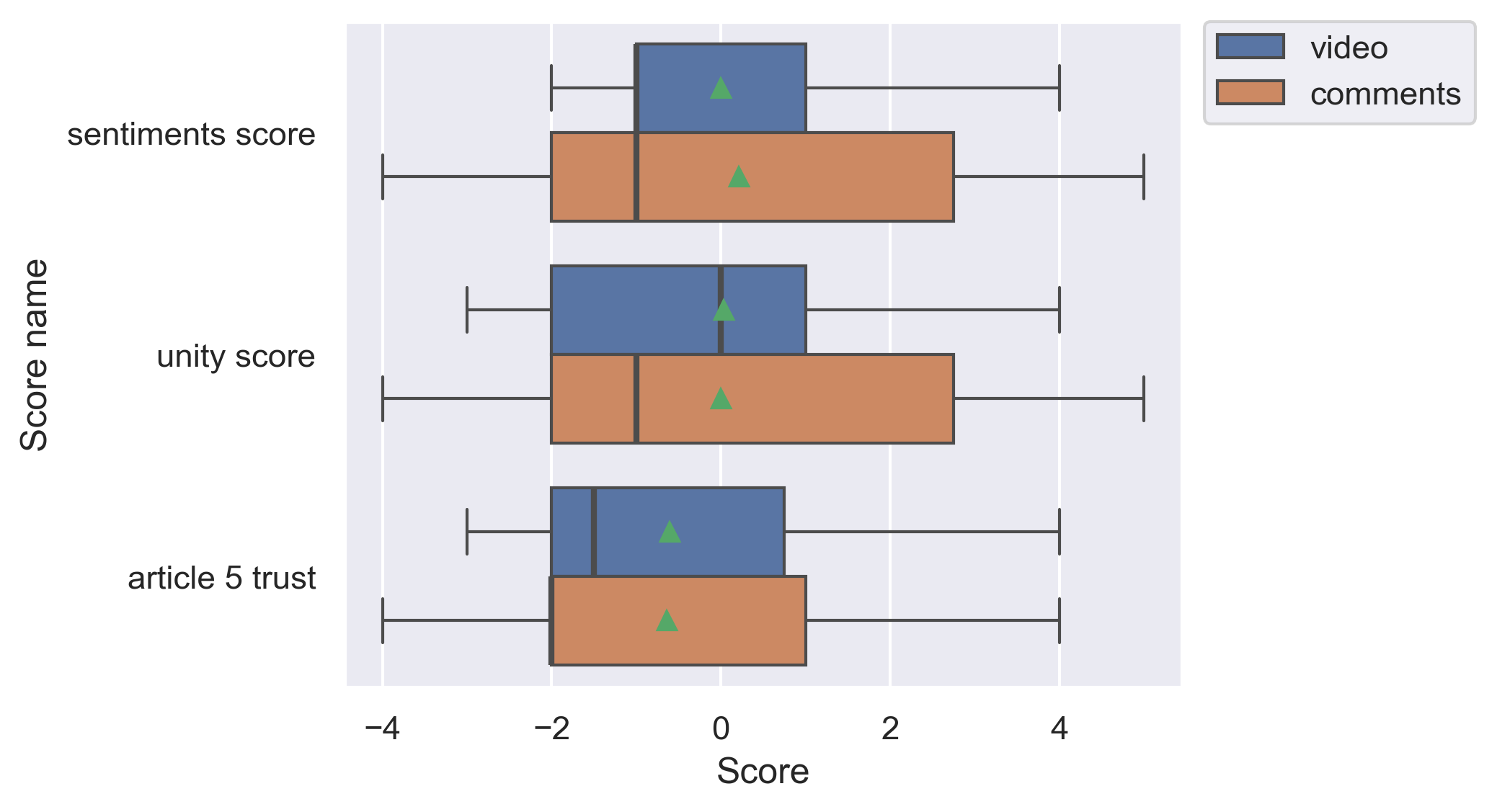}
\caption{Box plot for opinion scores on YouTube videos and their comments (triangles show mean values)}
\label{youtube_boxplot}
\end{figure}

For YouTube analytics, we used the following search queries: \textit{'NATO unity', 'NATO security', 'NATO article 5'} for time period of
2025-03-01 -- '2025-04-10'. The text of videos with more than 500K views along with users' comments were loaded. There were 16 videos. 
For the analysis, GPT-4.1 API with separate prompts for video texts and comments were used. In the prompt, we specified the instructions to generate quantitative scores in the range from -5 to 5 for \textit{"NATO sentiments","NATO unity", "NATO article 5"}  public opinions. 
Fig.~\ref{youtube_boxplot} presents the box plot for these scores. We can see that some scores have different distributions for videos and comments. 
On the second level of Youtube videos analytics,  mutual summarizing of video and their comments  was received using GPT-4.1 API. The resulting summary results are presented in the Appendix.

\FloatBarrier

\subsection{Reddit Discussions Analytics}
\begin{figure}[H]
\center
\includegraphics[width=0.7\linewidth]{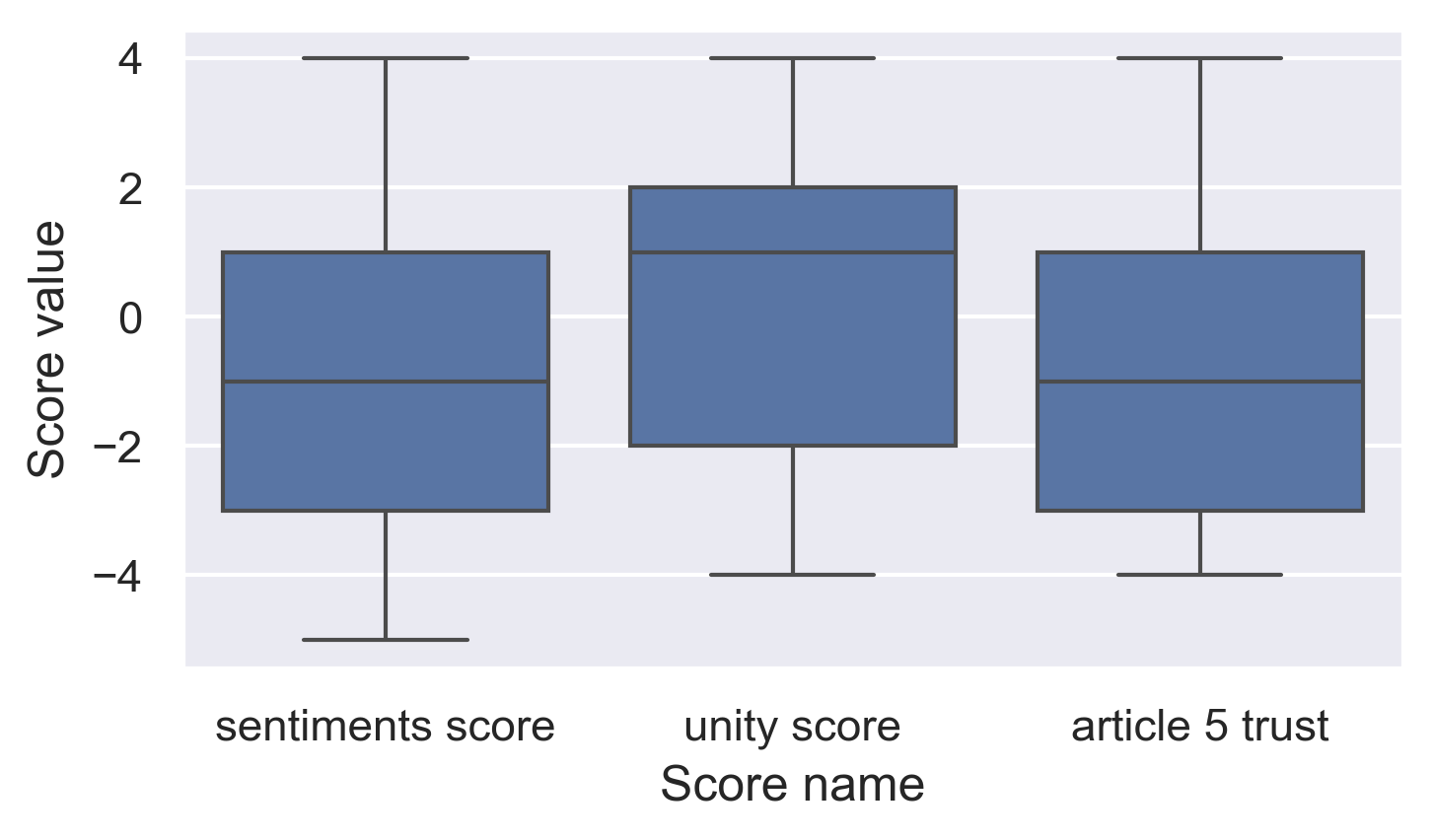}
\caption{Box plots for news opinion scores in Reddit discussions}
\label{reddit_scores}
\end{figure}

For the analysis, we loaded Reddit posts with discussion comments for the latest year using the  following queries: \textit{'NATO unity', 'NATO security', 'NATO article 5'} with the limit  100 for each query. 
Each post with the comment were summarized and opinion scores were generated using GPT-4.1 API. 
Fig.~\ref{reddit_scores} shows box plots for news opinion scores.

Summaries received on the second level of Reddit discussions analysis are presented in the Appendix.

\FloatBarrier
\section{Bayesian Regression for Opinion Scores}
\begin{figure}[H]
\center
\includegraphics[width=0.75\linewidth]{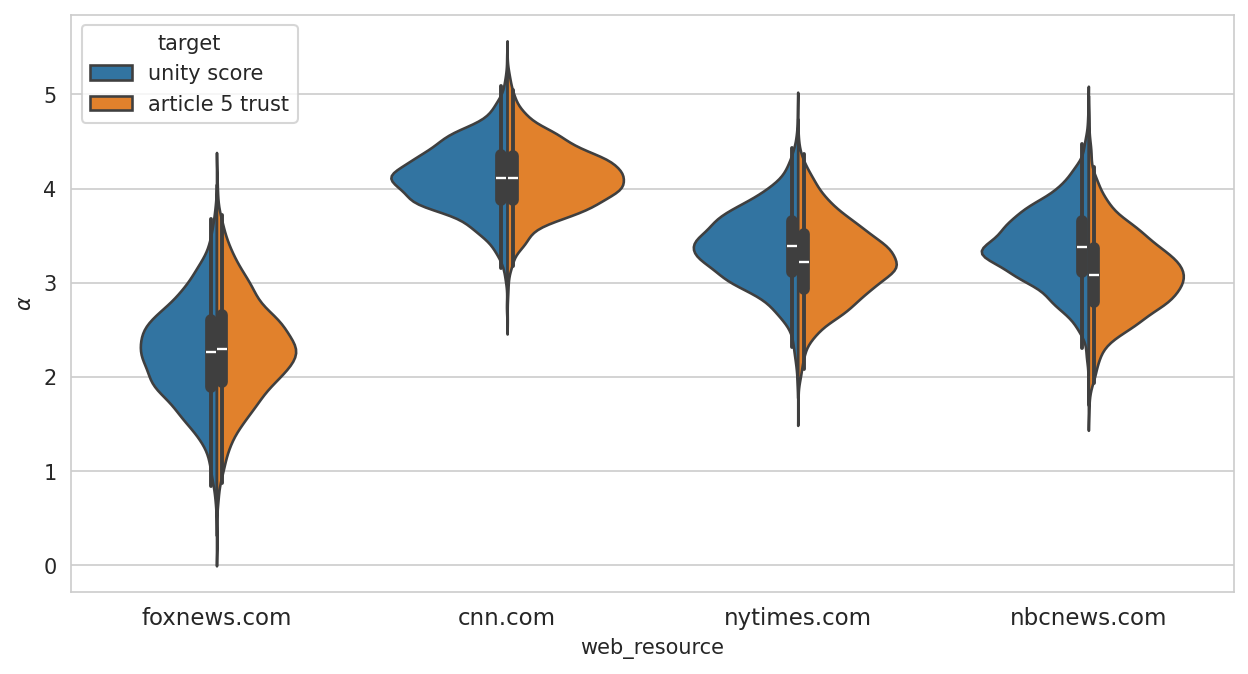}
\caption{ Box plot for $\alpha$ parameter of Bayesian regression for specified news resources}
\label{unity_score_alpha}
\end{figure}

\begin{figure}[H]
\center
\includegraphics[width=0.75\linewidth]{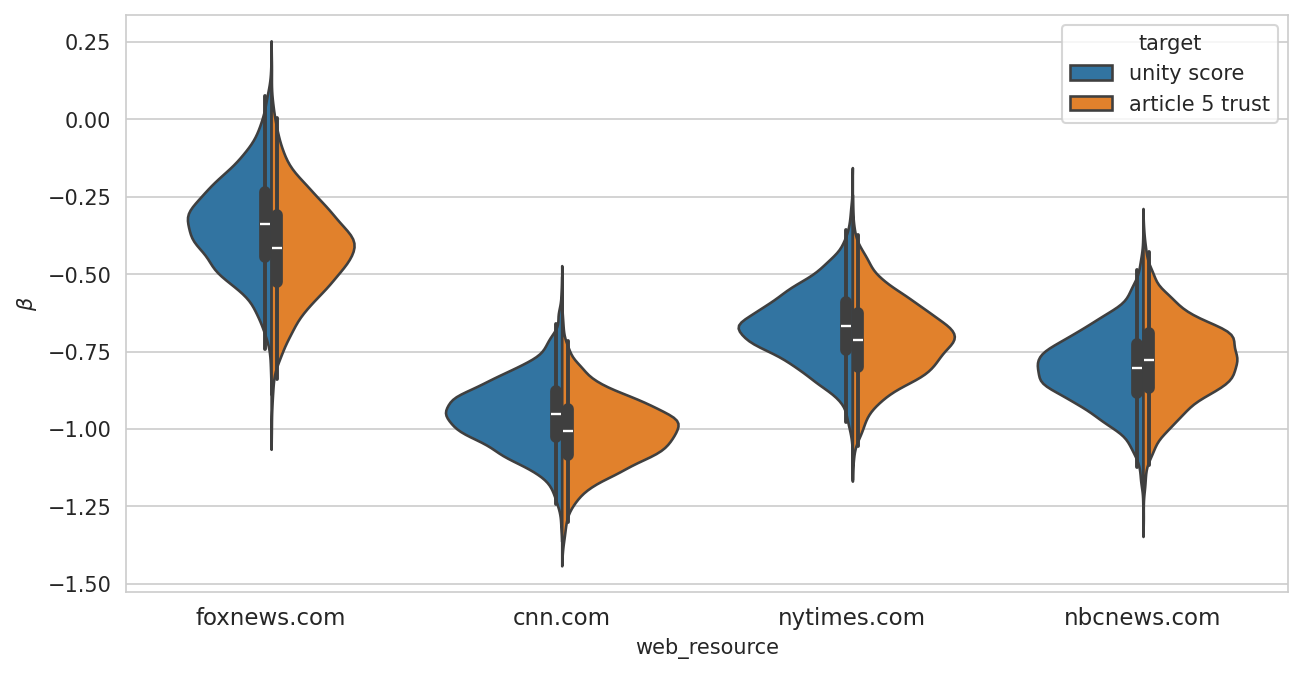}
\caption{Box plot for $\beta$ parameter of Bayesian regression for specified  news resources}
\label{unity_score_beta}
\end{figure}

To analyze trends, we can use Bayesian regression~\cite{kruschke2014doing,gelman2013bayesian,carpenter2017stan}.
 This approach allows us to receive a posterior distribution of model parameters by using conditional likelihood and prior distribution~\cite{pavlyshenko2020using, pavlyshenko2020regression}. 
 The probabilistic approach makes it possible to receive the probability density function for the target variable. 
To analyze trends for sentiment scores, we used the Bayesian regression model as follows:
\begin{equation}
 \begin{split}
 &Score \sim \mathcal{N}(\mu, \sigma) \\
 &\mu=\alpha + \beta t 
 \end{split}
\end{equation}
where $\alpha$ parameter describes the bias (intercept) of the trend, which can be treated as an initial score when $t=0$ in the time period under consideration, $\beta$ parameter describes the slope for upward or downward score trends.
To solve Bayesian models, numerical Monte-Carlo methods are used. Gibbs and Hamiltonian sampling are popular methods for finding posterior distributions of the parameters in probabilistic models
~\cite{kruschke2014doing,gelman2013bayesian, carpenter2017stan}.
Bayesian inference makes it possible to obtain probability density functions for model parameters and estimate the uncertainty which is important in risk assessment analytics. For Bayesian inference calculations, we used \textit{PyStan} package for \textit{Stan} platform for statistical modeling \cite{carpenter2017stan}. 
We conducted the  analysis for specified web sites: \textit{cnn.com,  nbcnews.com, nytimes.com, foxnews.com} and for NATO unity and Article 5 trust scores 
presented on Fig.~\ref{google_sites_img}.
For time independent variables, we used a list of indexes for time periods [1,2,3,4,5] which correspond to each annual time periods in range of 2021-2025.  
We conducted  Bayesian regression analysis for dynamic of opinion scores.  
 Boxplots for probability distributions of 
$\alpha$ and $\beta$ parameters for NATO unity and Article 5 trust scores are shown on Fig.~\ref{unity_score_alpha}, Fig.~\ref{unity_score_beta} respectively. The distributions found for the regression parameters make it possible to  analyse an uncertainty in the specified news opinion score trends. 

\section{Modeling Opinion Dynamics Using Neural Ordinary Differential Equations}
To analyse NATO unity opinion dynamics evolving, the approach of using  neural ordinary differential equations (ODE) can be used.
It can be useful when analysing different possible scenarious for opinion switching in the cases if they did not take place in previous time and as 
a result, we do not have historical data for analysis.
Different approaches for modelling social processes are described in~\cite{castellano2009statistical,galam2002minority,mobilia2003does,holme2006nonequilibrium,  sznajd2000opinion}.
Statistical physics has proven to be a fruitful framework to describe phenomena outside the realm of traditional physics. Recent years have witnessed an attempt by physicists to study collective phenomena emerging from the interactions of individuals as elementary units in social structures~\cite{castellano2009statistical}. 
The dynamics of spreading of the minority opinion in public debates (a reform proposal, a behavior change, a military retaliation) is studied using a diffusion reaction model. People move by discrete step on a landscape of random geometry shaped by social life (offices, houses, bars, and restaurants). A perfect world is considered with no advantage to the minority. A one person-one argument principle is applied to determine locally 
individual mind changes~\cite{galam2002minority}. 
A method for studying the exact properties of a class of inhomogeneous stochastic many-body systems is developed and presented in the framework of a voter model perturbed by the presence of a “zealot,” an individual allowed to favor an “opinion”~\cite{mobilia2003does}. 
Models of the convergence of opinion in social systems have been the subject of considerable recent attention in the physics literature. These models are divided into two classes, those in which individuals form their beliefs based on the opinions of their neighbors in a social network of personal acquaintances, and those in which, conversely, network connections form between individuals of similar beliefs. While both of these processes can give rise to realistic levels of agreement between acquaintances, practical experience suggests that opinion formation in the real world is not a result of one process or the other, but a combination of the two~\cite{holme2006nonequilibrium}. 
A simple Ising spin model which can describe a mechanism of making a decision in a closed community is proposed. It is shown via standard Monte Carlo simulations that very simple rules lead to rather complicated dynamics and to a power law in the decision time distribution. It is found that a closed community has to evolve either to a dictatorship or a stalemate state (inability to take any common decision). A common decision can be taken in a "democratic way" only by an open community~\cite{sznajd2000opinion}.

 Let us consider a dynamic model with two stable conditions which stand for positive and negative opinions 
on some processes in the society e.g. on NATO unity. 
The opinion dynamics model describes how individual or collective opinions evolve under the influence of internal feedback (amplification and resistance) and external influence. Internal feedback tends to stabilize moderate opinions but amplify strong ones, while external influences temporarily or permanently shift opinions, depending on their intensity and persistence.
It combines internal opinion dynamics (nonlinear) with external forcing (impulses).
We consider simple non-linear model which take into account external impulses which can describe different important public events or news.
We also include into the model the memory which describe evolving of external influence. 
The model can be written as follows:

\begin{equation}
\begin{aligned}
\frac{dx}{dt} &= a  x (1 - x^2) - b  x + c  E \\
\frac{dE}{dt} &= -d  E + I(t)
\end{aligned}
\label{ode_eq}
\end{equation}
where
$x(t)$ is the opinion at time, 
$t$,
$E(t)$ is an external influence (like media, external shocks) at time 
$I(t)$ is a time-dependent external impulse signal which can be social events or important news. 
Main variables and parameters of the model are the following:
\begin{itemize}
\item $x(t)$: Opinion state of a person or a group at time $t$.
Role: Main variable describing the current opinion or belief strength.
\item $E(t)$: External influence (e.g., news, propaganda, events).
Role: Mediates the effect of external factors on opinions.
\item $a$: Nonlinear feedback strength.
Role: Controls self-reinforcing opinion dynamics (amplifies strong opinions).
\item $b$: Internal damping.
Role: Models resistance to maintaining extreme opinions (makes opinions decay toward neutrality).
\item $c$: Coupling between external influence and opinion.
Role: Measures how much the external factor $E(t)$ shifts opinions.
\item $d$: Decay rate of external influence.
Role: Models how quickly external events' effects fade over time.
\item $I(t)$: Impulse signal which describe external influence. 
Role: Represents sudden events that temporarily change $E(t)$ (e.g., major news, crises).
\end{itemize}

Opinion $x(t)$ evolves based on
its own nonlinear self-reinforcement described by term $ax(1-x^2)$,
its internal damping described by term $-bx$, pulling 
$x$ back toward neutrality (e.g., moderate views) and 
external influence described by term $cE$,
direct "push" from media/events on opinion.
External factor $E(t)$ evolves according to
natural decay
described by term  $-dE$,
without new impulses, the influence fades over time (forgetting events).
The term  $(1-x^2)$ in the model provides nonlinear saturation,  strong opinions tend to stabilize around values of ±1. 
We can consider equations system~(\ref{ode_eq}) as Neural Ordinary Differential Equations (Neural ODEs)~\cite{chen2018neuralode}.
 The neural network serves as a universal function approximator for the right-hand side of the differential equation. This allows the model to learn complex, time-dependent dynamics directly from data, without requiring an explicit analytical form of the governing equations. During forward integration, the neural network defines the instantaneous rate of change of the system's state, which is then numerically integrated using standard ODE solvers to obtain the system's trajectory over time.
For the inverse problem, which involves identifying the optimal parameters of the neural network, gradient-based optimization methods are employed. In this context, backpropagation is used to compute gradients of a loss function with respect to the model parameters. 
The neural network provides a flexible and data-driven approximation of system dynamics in the forward pass, while backpropagation facilitates the learning of parameters by minimizing the discrepancy between predicted and observed states.
The foundational theory  of Neural Ordinary Differential Equations (Neural ODEs) is developed in~\cite{chen2018neuralode,chen2021eventfn}.
For numerical modelling, package  \textit{torchdiffeq}~\cite{torchdiffeq} was used.
Method \textit{torchdiffeq.odeint} solves ODEs by numerically stepping forward in time using adaptive or fixed-step methods (like Runge-Kutta), maintaining full PyTorch tensor support and allowing gradients to flow through the integration for optimization tasks.

\begin{figure}[H]
\center
\includegraphics[width=1\linewidth]{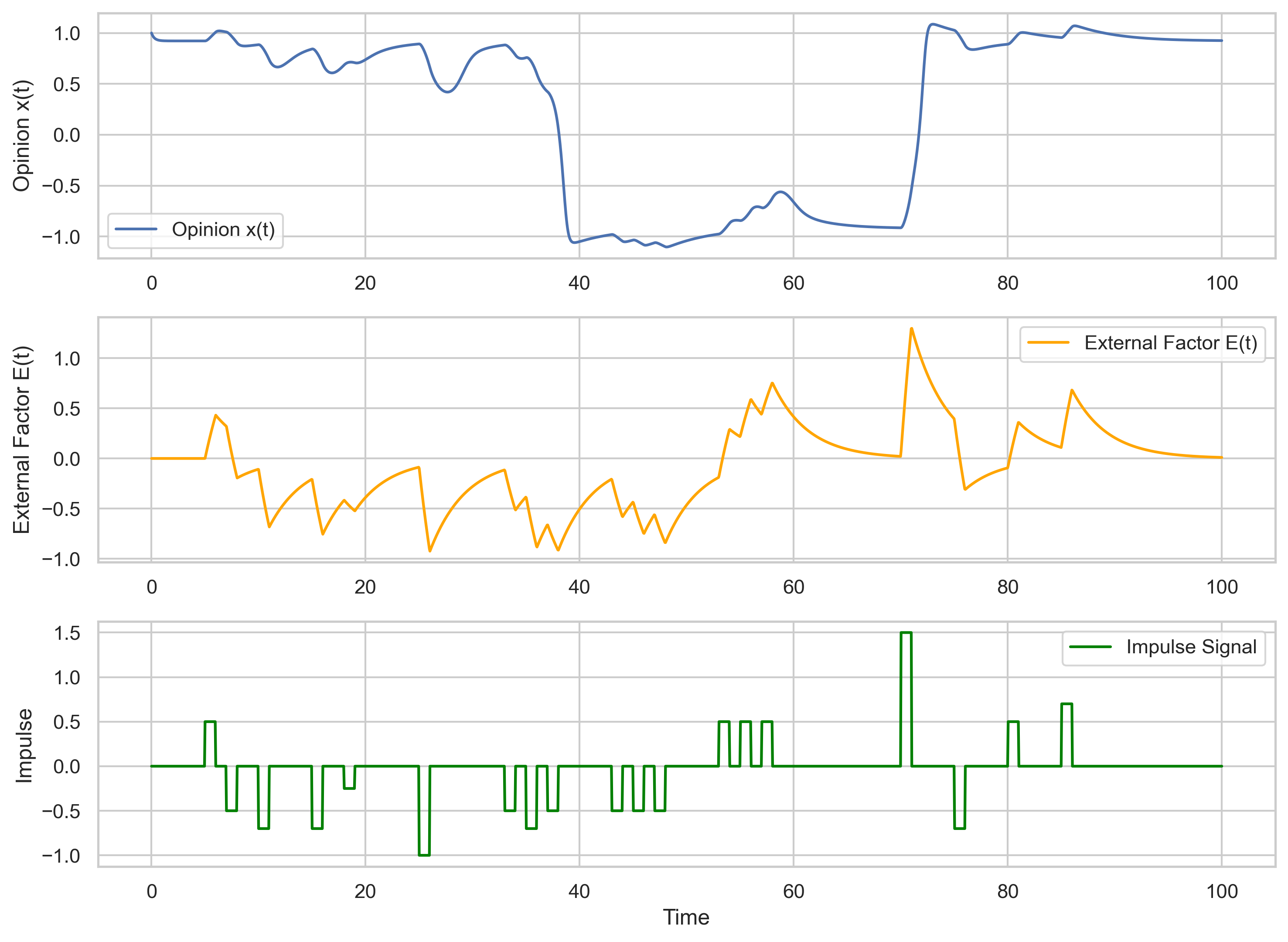}
\caption{Opinion dynamics with exogenous impulse}
\label{ode_modelling}
\end{figure}

For modeling we chose the following values for model~(\ref{ode_eq}) parameters:
\textit{a=2.0, b=0.3, c=1.0 , d=0.3}.
The obtained results show two switches in public opinion: the first one was caused by a group of influences, the second was caused by one exogenous impulse with large amplitude.
Fig.~\ref{ode_modelling} shows the results of numeric integration of the model using neural network algorithm. 
The results shows show that the opinion sign can be switched by a group of sequential impulses or by one sufficiently large impulse. Smaller impulses have influence on opinion value but do not switch opinion to opposite.

\begin{figure}[H]
\center
\includegraphics[width=0.7\linewidth]{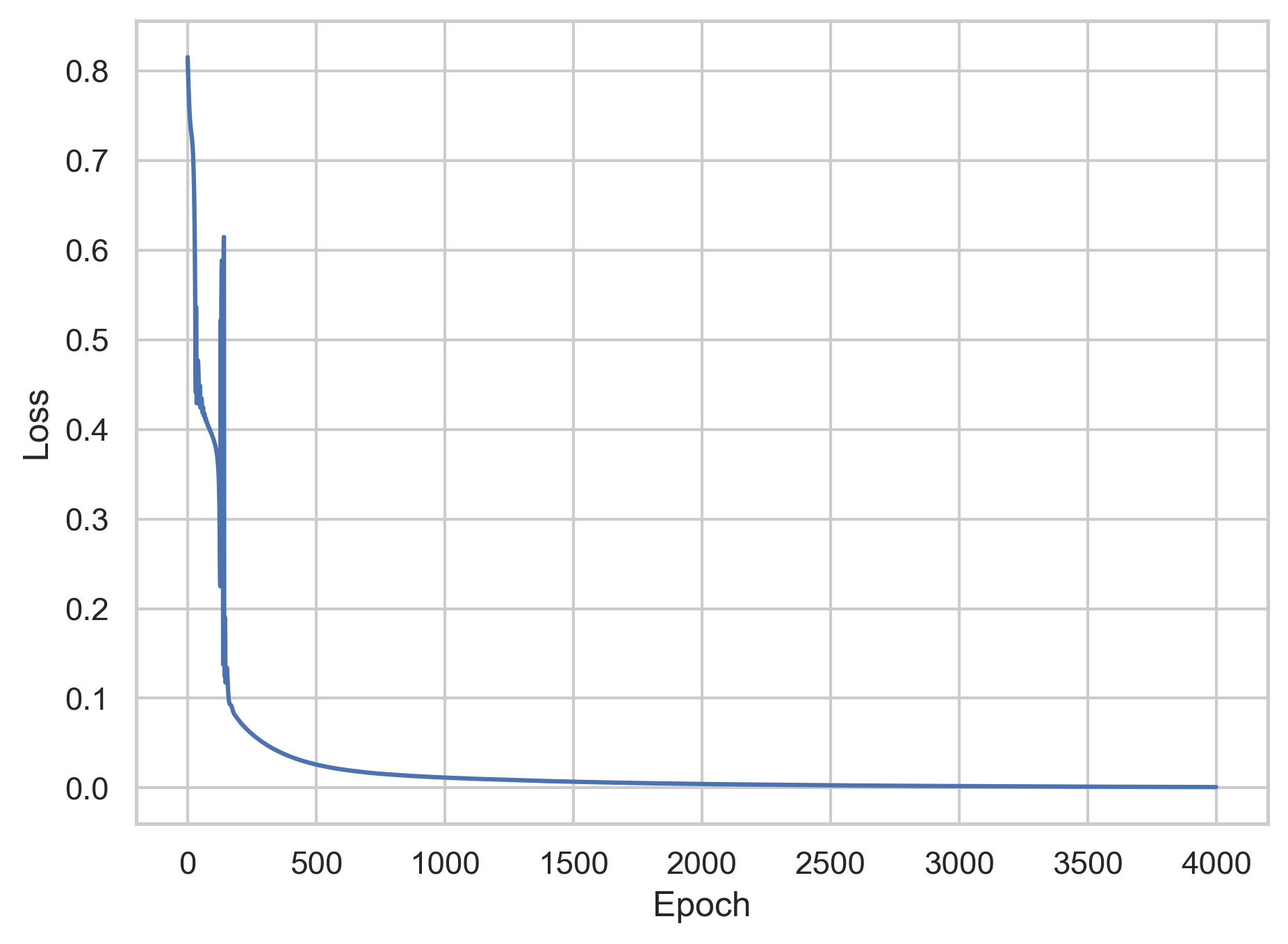}
\caption{Training loss}
\label{ode_loss}
\end{figure}

\begin{figure}[H]
\center
\includegraphics[width=0.7\linewidth]{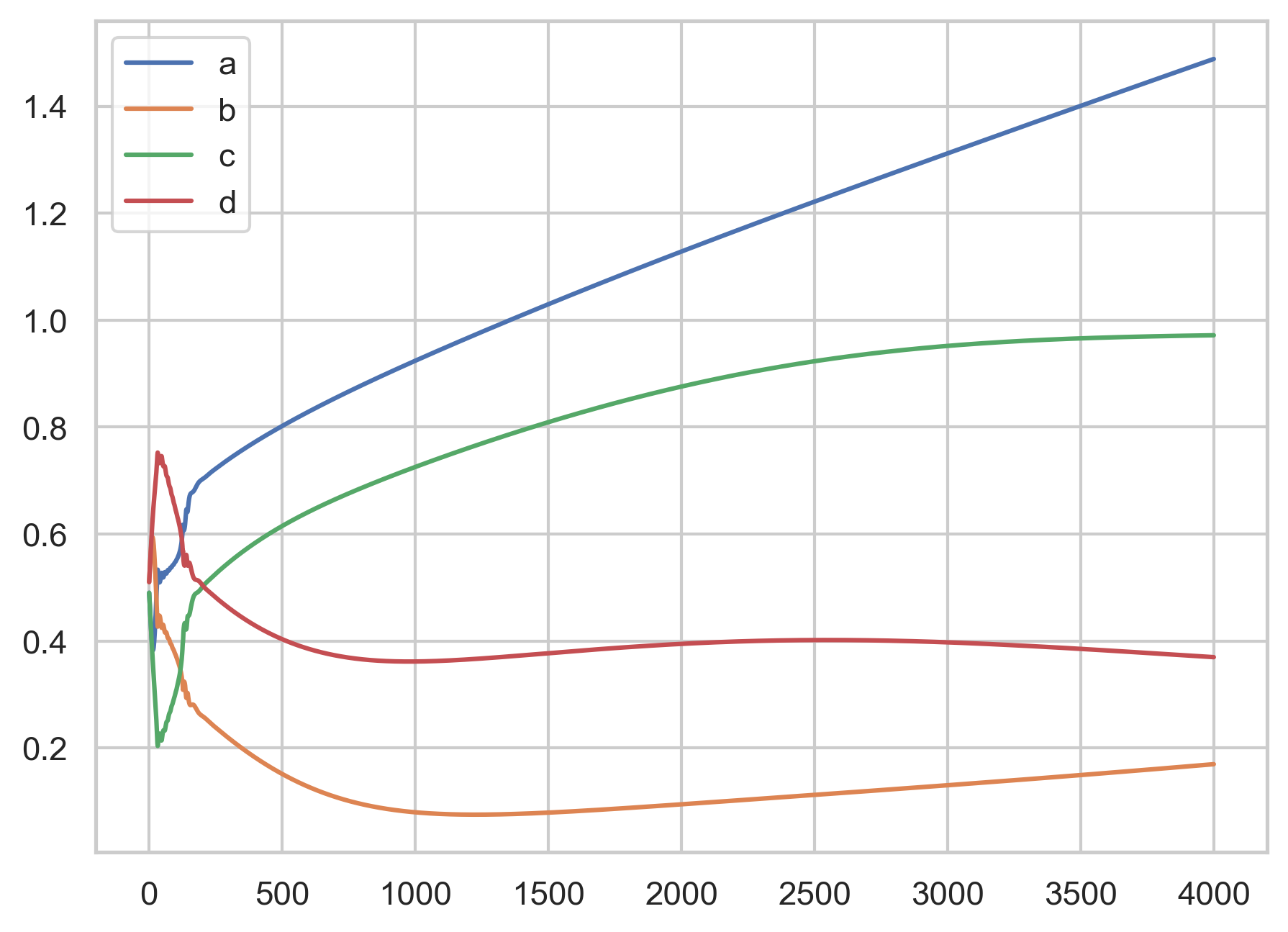}
\caption{Iterative optimization of model parameters in inverse problem}
\label{ode_reverse_par}
\end{figure}

It is important to consider inverse problem when we have experimental time series of opinion and need to find model parameters.It can be done using back propagation approach with libraries \textit{torchdiffeq}~\cite{torchdiffeq} and \textit{PyTorch}.
Fig.~\ref{ode_loss} shows the loss over training iterations, Fig.~\ref{ode_reverse_par} shows the dynamics of model parameters convergence during training iterations.
The model under consideration is very simple but it demonstrates the approach for modelling public opinion dynamics under external influences such as important social and  political events or news. 

\section{Conclusions}
In the work, we have considered the use  of GPT models with retrieval-augmented generation (RAG) for qualitative and quantitative analytics on NATO sentiments, NATO unity and NATO Article 5 trust opinion scores in different web sources:  news sites found via Google Search API, Youtube videos with comments and Reddit discusions.  The RAG approach using GPT-4.1 model  was applied to analyse  news where NATO related topics were discussed.   
Two-level RAG analytics were used: on the first level, the GPT model generates qualitative news 
summaries and quantitative opinion scores using zero-shot prompts; on the second level, the GPT model generates the summary of news summaries. 
Quantitative news opinion scores generated by the GPT model were analysed using Bayesian regression to get trend lines. The distributions found for the regression parameters make it possible to  analyse an uncertainty in the specified news opinion score trends.
The obtained results show a downward trend for analysed scores of opinion related to  NATO unity.
 The approach does not aim to conduct real political analysis; rather, it considers AI based approaches which can be used for further analytics 
 as a part of a complex analytical approach. The obtained results demonstrate that the use of GPT models for news analysis  can give informative qualitative and quantitative analytics, providing important insights. 
The dynamic model based neural ordinary differential equations were considered for modelling public opinion. The dynamic model based on neural ordinary differential equations was considered for modelling public opinions. This approach  makes it possible to  analyse different scenarios for evolving public opinions.
\section{Disclaimer} 
The approach, ideas, and results shared in this study are for academic purposes only and are not intended to inform real-world conclusions or recommendations.

\bibliographystyle{unsrt}
\bibliography{article.bib}

\begin{thebibliography}{10}

\bibitem{tardy2022nato}
Thierry Tardy.
\newblock {\em {" NATO 2030. United for a New Era": a Digest}}.
\newblock JSTOR, 2022.

\bibitem{kramer2009myth}
Mark Kramer.
\newblock {The myth of a no-NATO-enlargement pledge to Russia}.
\newblock {\em The Washington Quarterly}, 32(2):39--61, 2009.

\bibitem{lipka2019london}
R~Lipka.
\newblock {The London Summit as a Test for NATO Unity}.
\newblock {\em PULASKI POLICY PAPER. Assessed February}, 7:2021, 2019.

\bibitem{pavlyshenko2022methods}
Bohdan~M. Pavlyshenko.
\newblock {Methods of Informational Trends Analytics and Fake News Detection on
  Twitter}.
\newblock {\em arXiv preprint arXiv:2204.04891, Download PDF:
  https://arxiv.org/pdf/2204.04891.pdf}, 2022.

\bibitem{pavlyshenko2023analysis}
Bohdan~M Pavlyshenko.
\newblock {Analysis of Disinformation and Fake News Detection Using Fine-Tuned
  Large Language Model}.
\newblock {\em arXiv preprint arXiv:2309.04704}, 2023.

\bibitem{langchaingithub}
{LangChain}.
\newblock \url{https://github.com/langchain-ai/langchain}.

\bibitem{kruschke2014doing}
John Kruschke.
\newblock {\em {Doing Bayesian data analysis: A tutorial with R, JAGS, and
  Stan}}.
\newblock Academic Press, 2014.

\bibitem{gelman2013bayesian}
Andrew Gelman, John~B Carlin, Hal~S Stern, David~B Dunson, Aki Vehtari, and
  Donald~B Rubin.
\newblock {\em Bayesian data analysis}.
\newblock Chapman and Hall/CRC, 2013.

\bibitem{carpenter2017stan}
Bob Carpenter, Andrew Gelman, Matthew~D Hoffman, Daniel Lee, Ben Goodrich,
  Michael Betancourt, Marcus Brubaker, Jiqiang Guo, Peter Li, and Allen
  Riddell.
\newblock {Stan: A probabilistic programming language}.
\newblock {\em Journal of statistical software}, 76(1), 2017.

\bibitem{pavlyshenko2020using}
Bohdan~M. Pavlyshenko.
\newblock {Using Bayesian regression for stacking time series predictive
  models}.
\newblock In {\em 2020 IEEE Third International Conference on Data Stream
  Mining \& Processing (DSMP). Download PDF: https://arxiv.org/pdf/2201.02034},
  pages 305--309. IEEE, 2020.

\bibitem{pavlyshenko2020regression}
Bohdan~M. Pavlyshenko.
\newblock {Regression approach for modeling COVID-19 spread and its impact on
  stock market}.
\newblock {\em arXiv preprint arXiv:2004.01489. Download PDF:
  https://arxiv.org/pdf/2004.01489}, 2020.

\bibitem{castellano2009statistical}
Claudio Castellano, Santo Fortunato, and Vittorio Loreto.
\newblock {Statistical physics of social dynamics}.
\newblock {\em Reviews of modern physics}, 81(2):591--646, 2009.

\bibitem{galam2002minority}
Serge Galam.
\newblock {Minority opinion spreading in random geometry}.
\newblock {\em The European Physical Journal B-Condensed Matter and Complex
  Systems}, 25:403--406, 2002.

\bibitem{mobilia2003does}
Mauro Mobilia.
\newblock {Does a single zealot affect an infinite group of voters?}
\newblock {\em Physical review letters}, 91(2):028701, 2003.

\bibitem{holme2006nonequilibrium}
Petter Holme and Mark~EJ Newman.
\newblock {Nonequilibrium phase transition in the coevolution of networks and
  opinions}.
\newblock {\em Physical Review E—Statistical, Nonlinear, and Soft Matter
  Physics}, 74(5):056108, 2006.

\bibitem{sznajd2000opinion}
Katarzyna Sznajd-Weron and Jozef Sznajd.
\newblock {Opinion evolution in closed community}.
\newblock {\em International Journal of Modern Physics C}, 11(06):1157--1165,
  2000.

\bibitem{chen2018neuralode}
Ricky T.~Q. Chen, Yulia Rubanova, Jesse Bettencourt, and David Duvenaud.
\newblock {Neural Ordinary Differential Equations}.
\newblock {\em Advances in Neural Information Processing Systems}, 2018.

\bibitem{chen2021eventfn}
Ricky T.~Q. Chen, Brandon Amos, and Maximilian Nickel.
\newblock {Learning Neural Event Functions for Ordinary Differential
  Equations}.
\newblock {\em International Conference on Learning Representations}, 2021.

\bibitem{torchdiffeq}
Ricky T.~Q. Chen.
\newblock torchdiffeq, 2018.

\end{thebibliography}
\FloatBarrier
\newpage
\appendix
\section{Appendix}

\subsection{Using Natural Language Queries for GPT Aanlytics}
We tested the GPT-4.1 prompt  with inserted 
natural language query ”Give top 10 web resources with links where Article 5’s credibility is
openly doubted”.
 The following results were received:
{\fontsize{9}{11}\selectfont \begin{verbatim}
Top 10 Web Resources (2021-2025) Where Article 5’s Credibility is Openly Doubted

1. Title: Trump’s threat: NATO unity collapses amid U.S. support for Russia, Article 5 in doubt
Summary: The text highlights a dramatic decline in NATO unity due to US President Trump's overt
support for Russia and his questioning of America's commitment to defending European allies under
Article 5. European leaders are now doubting the reliability of the US as NATO's anchor, leading to
calls for increased defense autonomy within Europe. The credibility of Article 5 has been
undermined, with fears that US military support could be minimal or symbolic rather than
substantive. Overall, the narrative conveys major challenges to NATO's collective security, unity,
and trust in American leadership.
Link: https://www.bbc.com/news/articles/c0l1w1w41xzo

2. Title: The end of NATO? Trump’s retreat and Article 5 doubts prompt European crisis
Summary: The text discusses growing divisions within NATO, particularly as the US under the Trump
administration signals a retreat from its traditional security commitments, including questioning
the reliability of Article 5. There is concern that US disengagement may leave Europe vulnerable to
Russian aggression and force European states to consider developing their own nuclear deterrent. The
reliability of US-led collective defense is eroding, legal and political barriers to European self-
defense persist, and debates about trust in NATO's core commitments are intensifying. The risk of a
'dormant NATO' and divergent national interests threaten Alliance unity and security.
Link: https://verfassungsblog.de/the-end-of-nato/

3. Title: How Trump’s Trumpeting Has Dealt a Blow to NATO’s Article 5 Credibility
Summary: The text discusses the erosion of NATO's unity and security due to actions and rhetoric
from President Trump, who has questioned the U.S. commitment to collective defense and demanded
increased military spending from allies. This has undermined trust in Article 5, NATO's cornerstone
principle of mutual defense, as European allies are unsure if the U.S. would fulfill its obligations
in case of an attack. The credibility of NATO's security guarantees is described as being 'on life
support', and the possibility of U.S. withdrawal or non-commitment threatens the alliance's
cohesion, making it a strategic win for adversaries like Russia.
Link: https://www.msnbc.com/opinion/msnbc-opinion/nato-article-5-collective-defense-trump-rcna196981

4. Title: Trump: No security promises or NATO for Ukraine - fueling Europe’s Article 5 anxiety
Summary: The text discusses President Trump's refusal to provide US security guarantees or support
NATO membership for Ukraine, putting responsibility for Ukrainian security on European allies. This
divergence from traditional US leadership in NATO signals strains within the alliance, as key
European members see US support as critical for collective defense. The US siding with Russia on a
UN resolution and emphasizing resource deals over security commitments undermines trust in NATO's
Article 5 and overall alliance unity.
Link: https://www.lemonde.fr/en/united-states/article/2025/02/26/trump-says-no-security-promises-or-
nato-for-ukraine_6738609_133.html

5. Title: “Article 5: The pillar of NATO undermined by Donald Trump”
Summary: The article discusses the current challenge to NATO unity and security, primarily stemming
from US President Donald Trump's contesting the credibility of Article 5, the alliance's mutual
defense clause. While Article 5’s importance and deterrent effect are underscored, trust in this
provision is being undermined. The necessity for unanimous agreement before invoking collective
defense is seen as a vulnerability, raising questions about reliability if key members, particularly
the US, cast doubt on their commitment.
Link: https://www.lemonde.fr/en/international/article/2025/03/09/article-5-the-pillar-of-nato-
undermined-by-donald-trump_6738973_4.html

6. Title: BBC: Trump’s doubts shake NATO unity and Article 5
Summary: The text details rising concerns about NATO unity, driven by perceived US disengagement and
the US President's warm rhetoric towards Russia. European leaders, especially in Germany, now openly
discuss reducing reliance on the US for security. Trump's questioning of NATO's obligations, attacks
on Article 5's credibility, and encouragement of higher European defense spending have shaken trust
in collective defense. The uncertainty forces Europe to consider greater autonomy. Overall, the
greatest challenges for NATO unity are US unpredictability, diminished trust in Article 5, and a
destabilized security environment due to the war in Ukraine.
Link: https://www.bbc.co.uk/news/articles/c0l1w1w41xzo

7. Title: CBS News: Trump raises fresh doubts about NATO treaty’s Article 5
Summary: The text highlights significant challenges to NATO unity and security stemming from
President Trump's public doubts about allied commitment to collective defense and threats to
withhold U.S. protection from NATO members not meeting increased defense spending thresholds. These
statements raise concerns about the credibility of Article 5, the alliance's security guarantee.
While some European leaders reaffirm loyalty and call for increased defense spending, overall trust
in mutual defense commitments is undermined, putting NATO unity and the effectiveness of Article 5
at risk.
Link: https://www.cbsnews.com/news/trump-nato-article-5-collective-defense-europe-doubt-us-treaty-
commitment/

8. Title: CNN: Trump signals to Putin he won’t defend “delinquent” allies - Article 5 threatened
Summary: The text discusses former President Trump's public statements suggesting he would not
support NATO's collective defense (Article 5) for member countries not meeting defense spending
targets, instead encouraging Russian aggression against them. This has caused alarm among NATO
leaders, who argue such rhetoric undermines alliance unity and security. There are deep concerns
about the reliability of the US commitment to NATO's core Article 5, posing a significant challenge
to alliance unity and trust.
Link: https://www.cnn.com/2024/02/10/politics/trump-russia-nato/index.html

9. Title: Cato Institute: NATO enlargement and the fracturing of Article 5 credibility
Summary: The text critically examines the impact of NATO enlargement on US national security,
challenging the consensus that expansion has strengthened NATO unity and security. Enlargement has
fostered divisions, encouraged member states to rely excessively on US guarantees (cheap riding),
and provided reckless regional policies. The future prospect of Ukrainian membership threatens
alliance unity, as the US is unwilling to commit fully to Ukrainian defense, thus undermining trust
in Article 5 and raising fundamental questions about NATO's security guarantees, particularly to new
or aspirant members.
Link: https://www.cato.org/commentary/has-nato-enlargement-enhanced-us-national-security

10. Title: Responsible Statecraft: White House, Congress, and constitutional doubts over Article 5
Summary: The text discusses debates within the U.S. about the interpretation of NATO's Article 5.
Senator Rand Paul's view that Article 5 does not obligate automatic U.S. military action if a NATO
ally is attacked, and instead requires Congressional approval, has raised concerns about US
commitment. This stance challenges NATO unity and security, revealing vulnerabilities in alliance
cohesion and trust in Article 5 guarantees.
Link: https://responsiblestatecraft.org/2023/06/22/natos-article-5-does-not-override-congresss-war-
powers/

\end{verbatim}
}

\subsection{Summary for Youtube Videos and Comments}
On the second level of Youtube videos analytics using GPT-4.1 API,  we received the following resulting mutual summary of videos and their comments: 
{\fontsize{9}{11}\selectfont \begin{verbatim}
### Summary of Summaries

#### Challenges for NATO Unity, Security, and Trust in Article 5

Across the analyzed periods, several persistent, evolving challenges threaten NATO’s unity,
security, and the reliability of its Article 5 mutual defense guarantee:

- **US Political Uncertainty:** Repeated political shifts in the US—most notably linked to President
Trump’s transactional approach and public questioning of Article 5—emerge as the single largest risk
to alliance cohesion and credibility. Concerns abound about whether the US would honor Article 5 in
the event of Russian aggression, especially toward Eastern European states.
- **Intra-European Divisions:** Disagreements among European members, marked by varied defense
spending, differences in responses to crises (notably Ukraine), and caution in escalating
commitments, add a layer of fragmentation. Not all European states are equally willing to confront
Russia or support full NATO membership for Ukraine.
- **Burden-Sharing and Capability Gaps:** Slow European adaptation in military investment and
reluctance to fully assume strategic responsibilities—historically dominated by US capabilities—fuel
doubts about Europe’s capacity to stand alone if US disengagement materializes.
- **Member State Autonomy:** Turkey, in particular, demonstrates both increased NATO alignment and
continued strategic autonomy, sometimes acting independently (e.g., potential troop deployments to
Ukraine), complicating alliance coherence and mutual defense expectations.
- **Public and Elite Distrust:** Recurrent skepticism—among elites, publics, and commenters
alike—about the practical reliability of Article 5. Past broken security guarantees (e.g., Budapest
Memorandum) are used as cautionary examples, reflecting a widespread fear that collective defense
may be conditional and politically expedient.
- **Disinformation and Misunderstanding:** External propaganda, internal misinformation, and lack of
public understanding of what Article 5 truly entails further weaken perceived unity and trust.

#### Time Trends: Unity, Sentiment & Article 5 Trust

- **Unity Scores:** Unity perceptions fluctuate sharply with US political developments. Early
periods, marked by Trump-era rhetoric and subsequent uncertainty, show notable decreases in both
video and comment summaries. Unity slightly recovers in periods marked by new accessions (e.g.,
Sweden/Finland) or when faced with acute external threats from Russia, but the overall trend remains
volatile and vulnerable to US political signals.

- **Sentiment Scores:** Sentiment among both videomakers and commenters oscillates between cautious
optimism (following collective actions or new members joining) and deep frustration (following US
brinkmanship or intra-European disputes). Peaks of optimism are observed when member solidarity is
affirmed—primarily after Sweden’s accession or visible, practical cooperation—but these are
consistently dampened by lingering fears of US disengagement.

- **Article 5 Trust Scores:** Trust in Article 5 shows a consistent downward trend amid repeated
questioning by US leaders and transactional framing of mutual defense promises. Brief upticks occur
during periods of practical alliance action or when new members express faith in NATO, but overall,
trust appears increasingly fragile, especially in comments referencing recent historical betrayals
(Ukraine, Crimea) and conditionality of responses.

#### Groups of News/Video Sources by NATO Unity & Article 5 Trust Opinions

- **Strong Pro-Unity, High Article 5 Trust:** These sources (and their audiences) focus on
successful NATO expansions (e.g., Sweden joining), emphasize interoperability, and highlight mutual
support, especially from Northern/Western Europe and North America.
- **Cautious/Conditional Trust, Defensive Unity:** Sources centered on US politics, French autonomy,
and EU “strategic autonomy” communicate skepticism—unity is achievable, but only if Europe does more
independent work. Trust in Article 5 is often portrayed as contingent on US political developments.
- **Highlighting Divisions, Low Article 5 Trust:** Outlets focusing on US-EU disagreements, Russia’s
war in Ukraine, and Turkey’s maverick behaviors emphasize alliance rifts, unreliability
(particularly from the US), and frequent operational or strategic misalignments.
- **Narratives From/On Turkey:** Both mainstream and Turkish sources highlight Turkey’s “independent
actor” role—sometimes praised, sometimes seen as divisive. Trust in Article 5 is discussed as both
necessary and ambiguous in light of Turkey’s past actions at odds with consensus.
- **Populist/East European Skeptics:** Many comments and some sources focused on Eastern European
concerns repeatedly wonder whether “the West” would actually defend against Russian aggression,
often invoking Budapest and Article 5 ambiguities.

#### Consistency Between Video and Comment Summaries

- **High Alignment:** Generally, video narratives and comment discussions reinforce each
other—themes of uncertainty, anxiety over US commitments, and frustration with NATO’s inertia appear
both in expert/production-level summaries and grassroots commentary.
- **Divergence Points:** Occasionally, comments may be more optimistic or show greater expressions
of solidarity than the original videos, particularly in response to new accessions (e.g., Sweden).
Conversely, comments sometimes contain deeper cynicism and bitterness over perceived betrayals or
lack of action (Budapest, Article 5 doubts), reflecting perhaps more immediate or emotional
reactions than the videos themselves.

#### Key Takeaways

- **US Reliability is the Central Axis:** Trust in NATO unity and Article 5 is fundamentally
anchored in US policy consistency. Any sign of US political volatility or transactionalism
undermines the alliance at all levels.
- **Europe is (Slowly) Awakening:** There is an increasing realization—both in expert circles and
public sentiment—that Europe must develop greater strategic autonomy, invest more in defense, and
prepare to act if US support wanes or disappears.
- **Article 5 is Symbolically Strong but Practically Fragile:** While Article 5 is the heart of
NATO’s deterrence, repeated historical failures (e.g., perceived betrayal of Ukraine’s security
guarantees) and political hesitation are eroding its real-world credibility.
- **Unity is Event-Driven:** External threats (Russian aggression, hybrid warfare) drive temporary
unity spikes, especially following visible, collaborative gestures. However, these are vulnerable to
internal division and political wind shifts.
- **Commenters Echo, Amplify, and Sometimes Exceed Video Concerns:** Grassroots discussions almost
universally emphasize anxiety about US commitment and call for European initiative—at times
expressing greater skepticism or hope than the source videos, depending on context.
- **Turkey is a Double-Edged Member:** Turkey exemplifies both the strengths and strains of the
alliance—its military contributions are important, but its independent actions and ambivalence
introduce recurring uncertainty.

### Conclusion

NATO’s unity, perceived security, and the credibility of Article 5 are persistently tested by a
complex set of external and internal challenges. Chief among these are US political volatility,
European military adaptation, and the unpredictable behavior of key members like Turkey. Alliance
cohesion is event-dependent and constantly recalibrated in response to the threat environment and
member-state politics. Both media analyses and public opinions are deeply sensitive to these
dynamics, suggesting that NATO’s future credibility hinges not only on formal declarations, but on
tangible, collective action and the reliability of its most powerful members.

\end{verbatim}
}

\subsection{GPT Summary for Reddit Discussions}
On the second level of Reddit discussions analysis using GPT-4.1 model API, the following summary has been received:
{\fontsize{9}{11}\selectfont \begin{verbatim}
### Summary of Reddit Sentiment and Discourse on NATO Unity, Security, and Article 5 Trust

**Overall Summary & Highlighted Challenges**

Reddit discussions about NATO's unity, security, and trust in Article 5 show a highly polarized
landscape, marked by recurring skepticism, anxiety, and, less frequently, optimism. The principal
challenges highlighted include:

1. **Erosion of Trust in US Leadership:** Many posts express fears that shifting US political
priorities—especially under administrations perceived as isolationist or antagonistic (notably
Trump's)—could dramatically weaken NATO’s cohesion. Unpredictable US leadership, threats of troop
withdrawals, and transactional approaches to alliances cause doubts over the reliability of American
commitments to Article 5, which forms the foundation of NATO’s collective defense.

2. **Internal Divisions and Political Instability:** Diverging national interests and political
polarization within NATO members (US, Hungary, Turkey, Germany) are repeatedly cited as sources of
friction, undermining common purpose. Populist or authoritarian-leaning governments are perceived as
incompatible with NATO’s foundational values, threatening its unity and collaborative security
posture.

3. **Reliance on the US/Technological Dependence:** The alliance’s strong military and technological
dependence on the US (e.g., F-35 control) worries many Europeans, fostering calls for increased
European military and defense independence. The perception that US technical or policy decisions can
paralyze European defense capabilities raises major security concerns and undermines trust.

4. **Information Warfare and External Subversion:** Disinformation campaigns, primarily from Russia,
China, and Iran, are seen as exploiting societal divisions within the alliance, eroding trust in
institutions, and weakening Western unity against external threats. Open discussion acknowledges
foreign efforts at manipulation, but also points out that internal weaknesses (polarization, lack of
decisive action) leave NATO vulnerable.

5. **Doubts about Article 5's Credibility:** Recurrent skepticism surfaces about whether Article 5
would be invoked vigorously in a real crisis—especially regarding new or border countries under
threat. Incidents involving Canada or the Baltics, as well as skepticism from countries like
Ireland, amplify anxieties that NATO’s mutual defense guarantee may not translate into real-world
action.

6. **Europe’s Need for Autonomy:** Across the spectrum, there’s a strong call for Europe to enhance
defense spending, develop autonomous military capabilities, and reduce reliance on the US. Support
for a European army or EU military structures is often framed as a necessary hedge against US
unpredictability.

7. **Issues of Enlargement and Inclusion:** Debates about the integration of new members, such as
Ukraine and Sweden, highlight conflicting priorities, burden-sharing, and fears of overextension.
Some view membership expansion as essential for European security, while others see it as a
complicating factor threatening alliance unity.

**Time Trends—Unity, Sentiment, and Article 5 Trust**

- **Unity Scores:** The majority of scores for NATO unity are negative or low-positive, indicating
persistent doubts. Positive spikes appear after significant gestures of solidarity (e.g.,
anniversaries, bipartisan legislation reaffirming support), but are often offset by discussions of
emerging threats, US leadership changes, or controversial member states.
- **Sentiment Scores:** These largely mirror unity scores, with negative or cautious tones
prevailing when discussing US politics, technological dependency, or information warfare. Sentiment
briefly trends upward in rare moments of consensus (e.g., celebrating Baltic independence from
Russia or affirming unity in response to crises).
- **Article 5 Trust:** Trust in Article 5 frequently dips into negative territory—especially in
conversations concerning US reliability, EU-US relations, and divisive political events. Optimism is
most notable among Eastern European and Baltic perspectives, or when specifically referencing
collective defense successes—but skepticism is widespread elsewhere.

**Groups by Opinion on NATO Unity and Article 5 Trust**

1. **Optimists (High Unity/Trust, Positive Sentiment):**
   - Celebrate milestones (Baltic states' anniversaries, Sweden and Finland joining NATO).
   - Emphasize the value of collective defense, see NATO actions (like support for Ukraine) as proof
of continued relevance.
   - Often located in frontline states or among politicians/experts pushing for deeper integration.

2. **Realists/Cautious Supporters (Mixed/Conditional Trust):**
   - Support NATO but stress the need for more defense spending, European autonomy, and readiness.
   - Trust Article 5, but only if complemented by proactive policies and visible commitments (e.g.,
bipartisan US legislation, increased German/Swedish defense spending).
   - Advocate for a balanced approach between US leadership and increased European responsibility.

3. **Skeptics (Low Unity/Trust, Negative Sentiment):**
   - Distrust the US (especially under Trump or perceived isolationist leaders).
   - Question whether Article 5 will be reliably honored, particularly for countries on the
strategic periphery.
   - Cite perceived US betrayal, strategic ambiguity, or failures to protect Ukraine, Canada, or the
Baltics as evidence.
   - Often call for European or national strategic autonomy and criticize missed opportunities for
alliance reform.

4. **Disillusioned/Disengaged (Very Low/Mixed Unity and Trust):**
   - View NATO as a declining or obsolete alliance, often proposing alternative security structures
(European army, pan-European solutions).
   - Predominantly negative on US involvement, sometimes skeptical about even the viability of
Article 5.
   - Cohort includes voices from countries with traditions of neutrality (Ireland), left-leaning or
isolationist, and cynical voices frustrated with global politics.

5. **External Criticism/Anti-NATO Narratives:**
   - Some discussions are shaped by visible foreign propaganda or point out information warfare
targeting NATO unity.
   - These threads often generate internal debate, but the sheer presence of propaganda narratives
is noted as itself a security challenge to alliance cohesion.

**Conclusion**

Reddit’s aggregate discourse paints a picture of an alliance at a crossroads: while optimism and
unity spike during coordinated actions and anniversaries, persistent anxieties about US reliability,
internal divisions, technological dependence, and the credibility of Article 5 repeatedly erode
trust. Calls for European strategic autonomy and military capacity are louder than ever. The
underlying trend, discernible through unity, sentiment, and Article 5 trust scores, shows a
cautious, self-critical community—one engaged in open debate about the necessity of deep reforms to
secure NATO’s future relevance and reliability.

\end{verbatim}
}

\end{document}